\documentclass[fleqn,twoside]{article}
\usepackage{espcrc2}
\usepackage{graphicx}

\title{Heavy quark expansion parameters from lattice NRQCD
\thanks{presented by N.~Tsutsui}}

\author{JLQCD Collaboration:
        N.~Tsutsui\address{High Energy Accelerator Research
        Organization(KEK), Tsukuba, Ibaraki 305-0801, Japan},
        S.~Aoki\address{Institute of Physics, University of Tsukuba,
        Tsukuba, Ibaraki 305-8571, Japan},
        R. Burkhalter\address{Center for Computational Physics,
        University of Tsukuba, Tsukuba, Ibaraki 305-8577, Japan},
        M.~Fukugita\address{Institute for Cosmic Ray Research,
        University of Tokyo, Kashiwa, Chiba 277-8582, Japan},
        S.~Hashimoto$^{\rm a}$,\\
        K-I.~Ishikawa$^{\rm c}$,
        N.~Ishizuka$^{\rm b,c}$,
        Y.~Iwasaki$^{\rm b,c}$,
        K.~Kanaya$^{\rm b,c}$,
        T.~Kaneko$^{\rm a}$,
        Y.~Kuramashi$^{\rm a}$,
        M.~Okawa$^{\rm a}$,\\
        T.~Onogi\address{Yukawa Institute for Theoretical Physics,
        Kyoto University, Kyoto 606-8502, Japan},
        S.~Tominaga$^{\rm c}$,
        A.~Ukawa$^{\rm b,c}$,
        N.~Yamada$^{\rm a}$,
        T.~Yoshi\'e$^{\rm b,c}$}

\begin{document}

\newlength{\minitwocolumn}
\setlength{\minitwocolumn}{0.5\textwidth}
\addtolength{\minitwocolumn}{-0.5\columnsep}
\begin{abstract}
  Using the lattice NRQCD action for heavy quark, we calculate
  the heavy quark expansion parameters $\mu_{\pi}^2$ and
  $\mu_G^2$ for heavy-light mesons and heavy-light-light baryons.
  The results are compared with the mass differences among heavy
  hadrons to test the validity of HQET relations on the lattice.

{\large              
\vspace*{-30em}      
\hfill KEK-CP-114    
\vspace*{30em}       
}                    
\end{abstract}

\maketitle
\section{Introduction}
In the calculation of inclusive decay rates of the heavy hadron,
the heavy quark expansion (HQE) technique is widely used.
At the order $1/m_Q^2$ of HQE two nonperturbative parameters
\begin{eqnarray}
  \mu_{\pi}^2(H_Q) & \equiv &
  \frac{1}{2M_{H_Q}}
  \left\langle H_Q \left| \bar{Q}(i\vec{D})^2 Q \right| H_Q \right\rangle,
  \label{eq:mu_pi}
  \\
  \mu_G^2(H_Q) & \equiv &
  \frac{1}{2M_{H_Q}}
  \left\langle H_Q \left| \bar{Q}\vec{\sigma}\cdot\vec{B} Q \right| H_Q
  \right\rangle,
  \label{eq:mu_G}
\end{eqnarray}
appear in the calculation.
$H_Q$ represents a heavy-light meson or heavy-light-light baryon
(for $b$ hadrons, 
$H_b$ = $B$, $B^*$, $\Lambda_b$, $\Sigma_b$, $\Sigma_b^*$).
For instance, the lifetime ratio of $b$ hadrons is given as
\cite{Neubert:1997we}
\begin{eqnarray}
  \label{eq:lifetime_ratio}
  \lefteqn{
    \frac{\tau(H_b^{(1)})}{\tau(H_b^{(2)})}
    = 1 + \frac{\mu_\pi^2(H_b^{(1)})-\mu_\pi^2(H_b^{(2)})}{2m_b^2}
    }
  \nonumber\\
  & &
  + c_G \frac{\mu_G^2(H_b^{(1)})-\mu_G^2(H_b^{(2)})}{m_b^2}
  + O(1/m_b^3),
\end{eqnarray}
with $c_G\simeq$ 1.2.
While $\mu_G^2$ may be evaluated from experimental values of hyperfine
splitting, the determination of $\mu_\pi^2$ requires some theoretical
inputs.
It should be noted that the parameters are defined in the static limit:
$m_Q\to\infty$.
For heavy-light meson, $\mu_{\pi,G}^2$ has been calculated
using the lattice version of
the Heavy Quark Effective Theory \cite{Gimenez:1997av}.


In this work we calculate $\mu_\pi^2$ and $\mu_G^2$ on the lattice
using the NRQCD action for heavy quark.
Although the individual matrix element suffers from large perturbative
uncertainty due to power divergence in the matching calculation,
their differences like $\mu_\pi^2(H_b^{(1)})-\mu_\pi^2(H_b^{(2)})$
are free from the uncertainty of the operator.
We calculate both $\mu_\pi^2(H_b^{(1)})-\mu_\pi^2(H_b^{(2)})$ and
$\mu_G^2(H_b^{(1)})-\mu_G^2(H_b^{(2)})$, and compare them with the
corresponding predictions for mass splittings.

\section{HQET mass formula}
The parameters $\mu_{\pi}^2$ and $\mu_G^2$ can be indirectly obtained
from hadron masses, using
\begin{equation}
  M_{H_Q}-m_Q=
  \overline{\Lambda}
  +\frac{-\mu_{\pi}^2-\mu_G^2}{2m_Q}
  +O\left(\frac{1}{m_Q^2}\right),
  \label{eq:mass_formula}
\end{equation}
where $\overline{\Lambda}$ is the residual energy difference
between $M_{H_Q}$ and $m_Q$ surviving in the infinite heavy quark
limit.
$\mu_{\pi}^2$ and $\mu_G^2$ appear in the correction terms of
$O(1/m_Q)$.
Therefore, by considering proper mass differences, certain combinations
of $\mu_{\pi}^2$ and $\mu_G^2$ can be extracted.

For example, a difference of $\mu_G^2$ can be obtained from the mass
splitting in a spin multiplet, because $\overline{\Lambda}$ and
$\mu_{\pi}^2$ have the same value. 
Also, the spin averaged mass $M_{\bar{B}}=(M_B+3M_B^*)/4$
does not depend on $\mu_G^2$, because $\mu_G^2$ is proportional to the
spin of the light degrees of the freedom and the sum of $\mu_G^2$ in the
spin multiplet vanishes.

\begin{figure}[t]
  \includegraphics[height=\minitwocolumn,width=5.2cm,angle=-90]{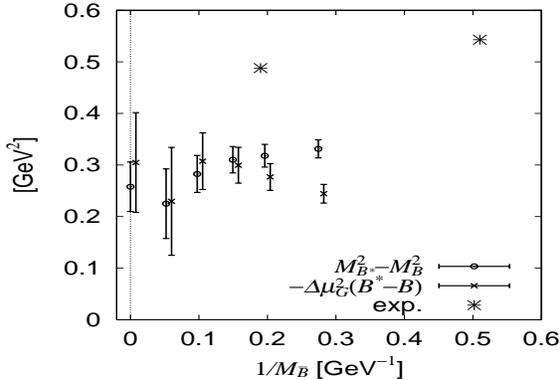}
  \vspace{-8mm}
  \caption{
    Hyperfine splitting of ground state mesons.
    Circles is obtained from the energy differences, while crosses are
    from the matrix elements.
    }
  \vspace{-2mm}
  \label{fig:hqdep.hfsm2.G.k0}
  \vspace{-5mm}
\end{figure}

\section{Lattice calculations}

We carry out quenched QCD simulations at $\beta$=6.0 on a
$20^3\times48$ lattice. 
The NRQCD action including all $O(1/m_Q)$ terms and the
non-perturbatively improved clover action ($c_{sw}$=1.769) 
is adapted for heavy quark and light quark, respectively.
Five heavy quark masses $am_Q$=1.3, 2.1, 3.0, 5.0, and 10.0 are used
to study the $1/m_Q$ dependence of hadron masses and matrix elements,
while three hopping parameters $K$=0.13331, 0.13384, and 0.13432
are simulated to extrapolate to the chiral limit $K_c$=0.135284(8).
The inverse lattice spacing $a^{-1}$=1.85(5) GeV is determined with
the $\rho$ meson mass $m_{\rho}$=770 MeV. 

We measure the three-point functions 
$\langle{\cal O}_{H_Q}(t)
 {\cal O}_{\pi,G}(t_{\cal O})
 {\cal O}_{H_Q}^\dagger(0)\rangle$,
where ${\cal O}_{H_Q}$ is an interpolating field to create or
annihilate the hadron $H_Q$, and 
${\cal O}_{\pi,G}$ is the operator to be measured,
$\bar{Q}(i\vec{D})^2Q$ or $\bar{Q}\vec{\sigma}\cdot\vec{B}Q$.
We divide them by 
$\langle{\cal O}_{H_Q}(t_1){\cal O}_{H_Q}^\dagger(0)\rangle$
to obtain the desired matrix elements $\mu_\pi^2$ and $\mu_G^2$.

\section{Hyperfine splittings}

From (\ref{eq:mass_formula}) the hyperfine splitting $M_{B^*}-M_B$ is
given by $-\Delta\mu_G^2/2m_Q$, or equivalently
\begin{equation}
  M_{B^*}^2-M_B^2 = -\Delta\mu_G^2 \equiv 
  -(\mu_G^2(B^*)-\mu_G^2(B)),
  \label{eq:hfsm2}
\end{equation}
at the leading order.
In Figure~\ref{fig:hqdep.hfsm2.G.k0}, we plot our results for
$-\Delta\mu_G^2$ together with the measurement of 
$M_{B^*}^2-M_B^2$.
We observe that the relation (\ref{eq:hfsm2}) is satisfied very well,
while both are significantly lower than the experimental values for
$B$ and $D$ mesons.

In deriving (\ref{eq:hfsm2}) we used a relation
\begin{equation}
  \Delta\mu_{\pi}^2=\mu_{\pi}^2(B^*)-\mu_{\pi}^2(B)=0,
  \label{eq:diffm}
\end{equation}
which holds in the static limit.
However, for the NRQCD action including the spin-magnetic interaction
term at $O(1/m_Q)$, the operator ${\cal O}_\pi$ mixes with 
${\cal O_G}$ at order $\alpha_s/m_Q$.
This is the reason why our result for $-\Delta\mu_G^2$ deviates
from that of the mass difference in the lighter heavy quark mass region.
In other words, the relation (\ref{eq:diffm}) may be considered as a
renormalization condition for the operator ${\cal O}_\pi$.

\begin{figure}[t]
  \includegraphics[height=\minitwocolumn,width=5.2cm,angle=-90]{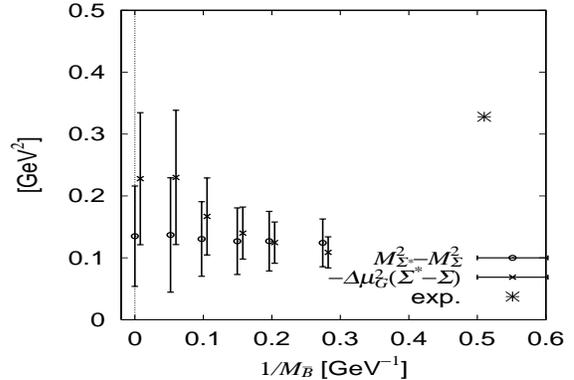}
  \vspace{-8mm}
  \caption{
    Hyperfine splitting of heavy-light-light baryons.
    }
  \vspace{-2mm}
  \label{fig:hqdep.hfsb2.G.k0}
\end{figure}

Similar analysis can be made for the hyperfine splitting of
heavy-light-light baryon, \textit{i.e.}
$\Sigma^*-\Sigma$ splitting.
Figure~\ref{fig:hqdep.hfsb2.G.k0} shows the mass difference and the
matrix element $-\Delta\mu_G^2$.
Both are in good agreement.

\section{$M_{\Lambda_b}-M_{\bar{B}}$}

\begin{figure}[t]
  \includegraphics[height=\minitwocolumn,width=5.2cm,angle=-90]{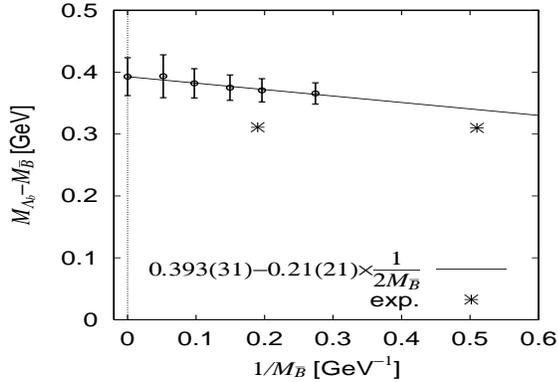}
  \vspace{-8mm}
  \caption{
    $1/M_{\bar{B}}$ dependence of $M_{\Lambda_b}-M_{\bar{B}}$.
    }
  \vspace{-2mm}
  \label{fig:hqdep.L-barB.mass}
\end{figure}

The heavy-light meson-baryon mass difference 
$M_{\Lambda_b}-M_{\bar{B}}$ is given as
\begin{eqnarray}
  M_{\Lambda_b}-M_{\bar{B}} &=&
  \overline{\Lambda}(\Lambda_b)-\overline{\Lambda}(B) \nonumber\\
  &+&\frac{1}{2m_Q}
  \left[-\mu_{\pi}^2(\Lambda_b)+\mu_{\pi}^2(B)\right].
  \label{eq:meson-baryon}
\end{eqnarray}
The intercept at $1/M_{\bar{B}}$=0 yields
$\overline{\Lambda}(\Lambda_b)-\overline{\Lambda}(B)$
while the slope is described by 
$-\Delta\mu_{\pi}^2=-(\mu_{\pi}^2(\Lambda_b)-\mu_{\pi}^2(B))$.

In Figure~\ref{fig:hqdep.L-barB.mass} we plot 
$M_{\Lambda_b}-M_{\bar{B}}$ as a function of $1/M_{\bar{B}}$.
For the intercept we obtain 
$\overline{\Lambda}(\Lambda_b)-\overline{\Lambda}(B)$=393(31)~MeV.
in agreement with a previous work by Ali~Khan \textit{et al.},
$\overline{\Lambda}(\Lambda_b)-\overline{\Lambda}(B)$=415(156)~MeV.
Our result is slightly larger than the experimental values for $b$ and
$c$ hadrons.
However, to draw a definite conclusion we have to consider several
systematic errors, especially the finite volume effect, because our
lattice may not be large enough for baryons.

\begin{figure}[t]
  \includegraphics[height=\minitwocolumn,width=5.2cm,angle=-90]{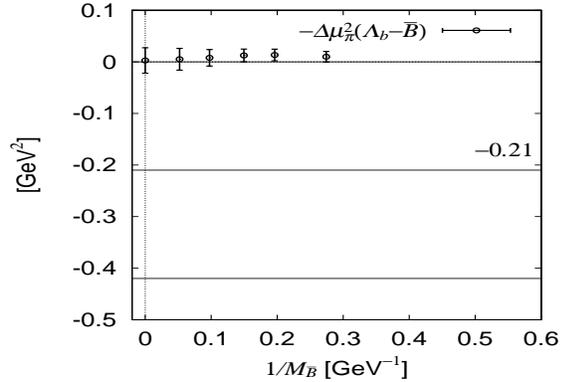}
  \vspace{-8mm}
  \caption{
    $-\Delta\mu_\pi^2$ measured from the matrix elements is compared
    with the slope of mass difference $M_{\Lambda_b}-M_{\bar{B}}$,
    $-0.21(21)$~GeV$^2$.
    }
  \vspace{-2mm}
  \label{fig:hqdep.L-barB.pi.k0}
\end{figure}

The slope obtained from the fit of the mass difference is consistent
with zero: $-0.21(21)$ GeV$^2$.
Our results of direct measurement of
$-\Delta\mu_\pi^2$ is plotted in Figure~\ref{fig:hqdep.L-barB.pi.k0},
which is consistent with the result from mass difference, but have much better
accuracy.
Our result is also compatible with the phenomenological estimate
$-0.01(3)$~GeV$^2$ \cite{Neubert:1997we}
obtained from a combination 
$(M_{\Lambda_b}-M_{\bar{B}})-(M_{\Lambda_c}-M_{\bar{D}})$.

\section{Conclusions}
We confirm that the lattice measurements of the matrix elements
$\mu_\pi^2$ and $\mu_G^2$ are consistent with the HQET mass relations.
The well-known problem of quenched lattice calculation that the
hyperfine splitting is much smaller than the experiments is also
reproduced.

An important extension of our work is to measure the matrix elements
of four-quark operators, which are relevant to the $1/m_Q^3$
corrections to the lifetime ratios \cite{Neubert:1997we}.

\vspace*{5mm}
This work is supported by the Supercomputer Project No.66 (FY2001)
of High Energy Accelerator Research Organization (KEK),
and also in part by the Grants-in-Aid of the Ministry of 
Education (Nos. 10640246, 11640294, 12014202, 12640253, 12640279,
12740133, 13640260 and 13740169).
K-I.I and N.Y are supported by the JSPS Research Fellowship.


\begin{thebibliography}{9}
\bibitem{Neubert:1997we}
  M.~Neubert and C.~T.~Sachrajda,
  Nucl.\ Phys.\ B {\bf 483} (1997) 339.

\bibitem{Gimenez:1997av}
  V.~Gimenez, G.~Martinelli and C.~T.~Sachrajda,
  Nucl.\ Phys.\ B {\bf 486} (1997) 227.

\bibitem{AliKhan:2000yb}
  A.~Ali~Khan {\it et al.},
  Phys.\ Rev.\ D {\bf 62} (2000) 054505.

\end{thebibliography}
\end{document}